# CONSUMER ACCEPTANCE OF THE USE OF ARTIFICIAL INTELLIGENCE IN ONLINE SHOPPING: EVIDENCE FROM HUNGARY

**Szabolcs Nagy[1*] and Noémi Hajdú[2]**
[1)2)] *University of Miskolc, Miskolc, Hungary*

**Abstract**
The rapid development of technology has drastically changed the way consumers do their shopping. The volume of global online commerce has significantly been increasing partly due to the recent COVID-19 crisis that has accelerated the expansion of e-commerce. A growing number of webshops integrate Artificial Intelligence (AI), state-of-the-art technology into their stores to improve customer experience, satisfaction and loyalty. However, little research has been done to verify the process of how consumers adopt and use AI-powered webshops. Using the technology acceptance model (TAM) as a theoretical background, this study addresses the question of trust and consumer acceptance of Artificial Intelligence in online retail. An online survey in Hungary was conducted to build a database of 439 respondents for this study. To analyse data, structural equation modelling (SEM) was used. After the respecification of the initial theoretical model, a nested model, which was also based on TAM, was developed and tested. The widely used TAM was found to be a suitable theoretical model for investigating consumer acceptance of the use of Artificial Intelligence in online shopping. Trust was found to be one of the key factors influencing consumer attitudes towards Artificial Intelligence. Perceived usefulness as the other key factor in attitudes and behavioural intention was found to be more important than the perceived ease of use. These findings offer valuable implications for webshop owners to increase customer acceptance.

* Corresponding author, **Szabolcs Nagy** – e-mail: nagy.szabolcs@uni-miskolc.hu

## Introduction

The rapid development of digital technology has changed online shopping (Daley, 2018). In recent years, the use of Artificial Intelligence (AI) in online commerce has been increased since AI is an excellent tool to meet rapidly changing consumer demand and to increase sales efficiency. The global spending by retailers on AI services is expected to quadruple and reach $12 billion by 2023, and over 325000 retailers will adopt AI technology (Maynard, 2019).

Smidt and Power (2020) claimed that online product research has significantly increased over the past years. USA's largest online retailer, Amazon, is the exemplary case of how to effectively integrate AI into online retail. Besides the rich assortment, fast delivery and competitive prices, a more localised shopping journey can be created. Thus Amazon can use location-specific pricing and send destination-specific messages to its customers, who will pay in their local currency (Barmada, 2020).

Novel marketing techniques supported by new technologies, including the use of AI systems spark the proliferation of new marketing methods to effectively reach target consumers and to offer enhanced consumer experiences (Pusztahelyi, 2020). Pursuant to Asling (2017), the use of AI in online shopping makes customer-centric search and a new level of personalisation possible resulting in a more efficient sales process. Information technology (IT) has changed the nature of company-customer relationships (Rust and Huang, 2014). However, any technology-driven transformation is based on trust (Pricewaterhouse Coopers, 2018).

Online retailers need more in-depth insight into how consumers perceive and accept the use of AI in webshops and how much they trust them. They also need to know how to use AI most effectively to increase online spending and online purchase frequency since the importance of time and cost efficiency in shopping has recently become more and more critical. In this regard, online shopping means a convenient way for customers to buy the desired products.

So far, only a few researchers have addressed the question of trust and consumer acceptance of AI in online retail. Based on the technology acceptance model (TAM), this study aims to fill this research gap and proposes an integrated theoretical framework of consumers' acceptance of AI-powered webshops. Further objectives of this paper are to investigate the relationships between the elements of TAM; to analyse the effects of trust, perceived usefulness and perceived ease of use on attitudes and behavioural intention.

After reviewing the use of AI in online shopping, this paper discusses the role of trust in online shopping and presents the technology acceptance model. The next section deals with the research methodology, including the research questions, hypotheses and the sample. In the results and discussion section, the validity and reliability of the model, as well as the model fit are presented. Hypothesis testing, detailed analysis of the relationships between the elements of the nested model, and comparison of the results with the previous research findings are also discussed here before the conclusions sections.

## 1. Literature review

According to IBM's U.S. Retail Index, the COVID-19 has speeded up the change from traditional shopping to online purchasing by circa five years (Haller, Lee and Cheung, 2020). Due to the pandemic situation, there is an increased demand for AI in the retail industry (Meticulous Market Research, 2020).

### 1.1. The use of AI in online shopping

AI systems are a set of software and hardware that can be used to continuously assess and analyse data to characterise environmental factors and to determine decisions and actions (European Commission, 2018). Prior research mainly focused on the advantages of the use of AI in online settings and failed to address how consumers accept AI in online retail. According to utility theory, this new technology helps consumers to find and choose the best product alternatives, while decreases the search cost and search time (Pricewaterhouse Coopers, 2018), thus increasing utility (Stigler, 1961; Bakos, 1977; Stigler and Becker, 1977; André, et al. 2017; Lynch and Ariely, 2000). AI filters the information for each target customer and provides what exactly is needed (Paschen, Wilson and Ferreira, 2020). AI supports automating business processes, gains insight through data analysis, and engages with customers and employees (Davenport and Ronanki, 2018).

Artificial intelligence is widely used to increase the efficiency of marketing (Kwong, Jiang, and Luo, 2016) and retail (Weber and Schütte, 2019) and to automate marketing (Dumitriu and Popescu, 2020). AI-powered online stores provide their customers with automated assistance during the consumer journey (Yoo, Lee and Park, 2010; Pantano and Pizzi, 2020). It is a great advantage, especially for the elder people, who are averse to technical innovations.

Consumers' online information search and product selection habits can be better understood by AI to offer a more personalised shopping route (Rust and Huang, 2014). It is a great opportunity for online shops to analyse the profile of existing and potential customers and thereby suggest tailor-made marketing offerings for them (Onete, Constantinescu and Filip, 2008). AI also makes the contact with both the customers and the employees continuous and interactive. Frequently asked questions (FAQs) regarding the products, product-use and ordering process can be automated by a chatbot. New sales models use automated algorithms to recommend unique, personalised marketing offerings, thus increasing customer satisfaction and engagement. To sum up the advantages, AI systems operate automatically and analyse big data in real-time to interpret and shape consumer behavioural patterns to offer products and services in a personalised way, thus enhancing the shopping experience.

However, AI systems also have some disadvantages. They work most effectively with big data; therefore, the implementation of AI systems requires huge investments (Roetzer, 2017).

### 1.2. The role of trust in online shopping

Trust is of great importance in online commerce. According to Kim, Ferrin and Rao (2008), consumer confidence has a positive effect on a consumer's intention to buy. The higher the consumer trust in an online shop is, the more likely the consumer will be to go through the buying process. Trust is especially crucial when the customer perceives a financial risk.

Thatcher et al. (2013) identified two types of trust: general and specific trust. General trust concerns the e-commerce environment, consumer beliefs about and attitudes towards it. Specific trust is related to the shopping experience in a specific virtual store. Confidence can be enhanced through interactive communication between the retailer and the buyer by using appropriate product descriptions and images to reduce the perceived risk. As stated in Cătoiu et al. (2014) there is a strong negative correlation between perceived risks and trust. According to Reichheld and Schefter (2000, p. 107), “price does not rule the Web; trust does”.

Aranyossy and Magisztrák (2016) found that a higher level of e-commerce trust was associated with more frequent online shopping. However, when shopping online, customers do not necessarily notice that a website uses AI tools (Daley, 2018).

All things considered, AI marks a new era in online sales. However, continuous technological development such as the use of AI-powered websites divides society, as there are those who accept novelty while others reject it.

### 1.3. Technology Acceptance Model (TAM)

Consumers' adaptation to new technologies can be explained by several models. Dhagarra, Goswami and Kumar (2020) summarised them as follows: (1) Theory of Reasoned Action (TRA) by Fishbein and Ajzen (1975); (2) Theory of Planned Behaviour (TPB) by Ajzen (1985); (3) Technology Acceptance Model (TAM) by Davis (1986); (4) Innovation Diffusion Theory (IDT) by Rajagopal (2002); (5) Technology Readiness Index (TRI) by Parasuraman, (2000); and (6) Unified Theory of Acceptance and Use of Technology (UTAUT) by Venkatesh, et al. (2003).

Technology acceptance model (TAM), an extension of (TRA), is one of the most widely-used theoretical models (Venkatesh, 2000) to explain why an IT user accepts or rejects information technology and to predict IT user behaviour (Legris, Ingham, and Collerette, 2003). The original TAM contains six elements: external variables, perceived usefulness, perceived ease of use, attitude, behavioural intention to use and actual use. According to TAM, external variables have a direct influence on perceived usefulness (PU) and perceived ease of use (PEU), i.e. the two cognitive belief components. Perceived ease of use directly influences PU and attitude, whereas perceived usefulness has a direct impact on attitude and behavioural intention to use, which affects actual use (Figure no. 1).

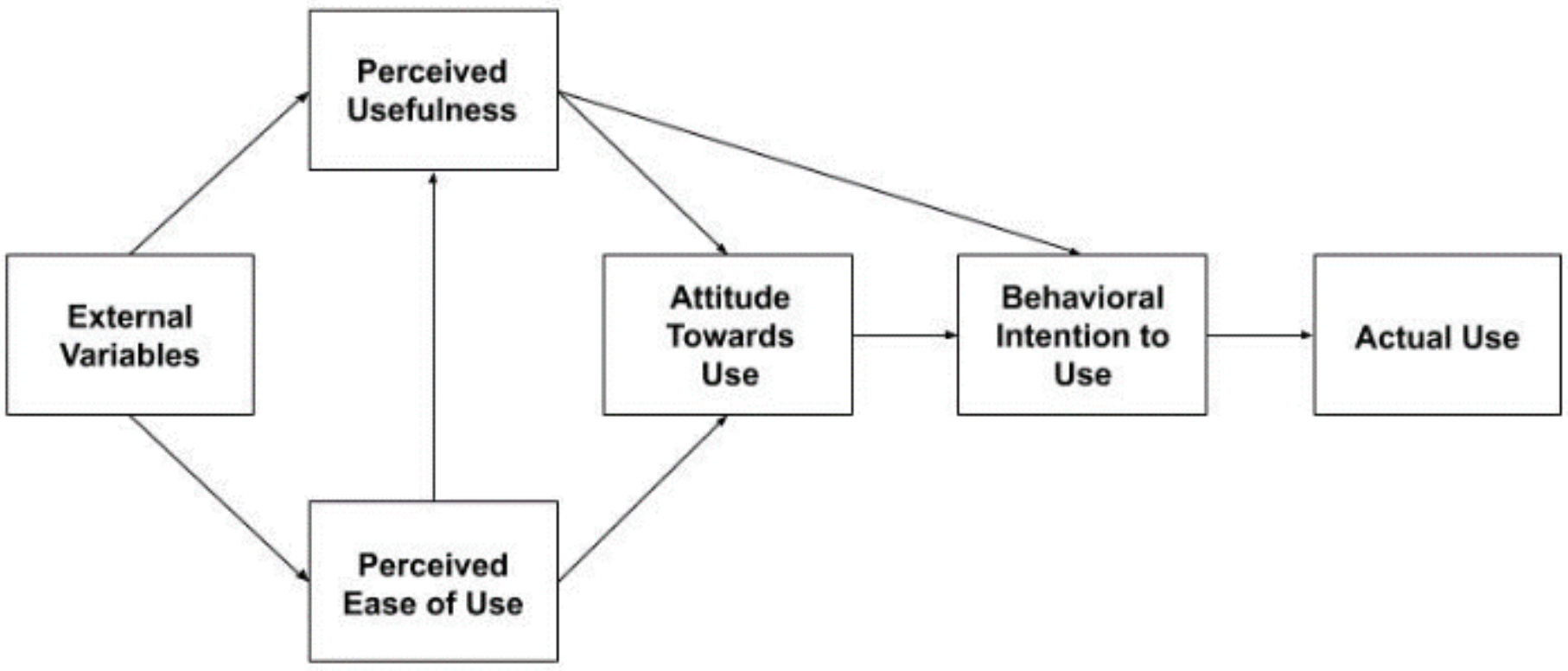

**Figure no. 1. The original technology acceptance model (TAM)**
*Source: Davis, 1986.*

Ha and Stoel (2008) examined the factors affecting customer acceptance of online shopping and found that perceived ease of use, perceived trust and perceived shopping enjoyment had the greatest impact on customer acceptance. Ease of use, trust and shopping enjoyment had a significant impact on perceived usefulness; trust, shopping enjoyment, and usefulness had a significant effect on attitude towards online shopping. They also found that attitude and perceived usefulness had an influential role in consumer intention to purchase online.

According to Vijayasarathy (2004), there is a positive association between consumer attitude towards online shopping and the beliefs concerning usefulness, compatibility, security and ease of use. Also, the intention to purchase online is strongly influenced by consumer beliefs about online shopping, self-efficacy and attitude. Surprisingly, no positive relationship between purchasing intention and consumer beliefs about the usefulness of online shopping was reported (Vijayasarathy, 2004). Gefen, Karahanna and Straub (2003) found that perceived usefulness and perceived ease of use influence consumer repurchase intention.

It must be noted that Schepman and Rodway (2020) expressed some criticisms about the applicability of TAM to measure attitudes towards AI. According to them, it is the online retailers that can decide to integrate AI into webshops, and consumers have no choice but to use it when shopping online in such stores. Therefore, traditional technology acceptance models might not be ideal to measure attitudes towards AI. However, we are convinced that consumers still have the free will to decide whether to use new technology, i.e. to shop online in an AI-powered webshop, or not.

## 2. Methodology and research questions

### 2.1. Methodology

The constructs and the measurement instruments presented in Table no. 1 were developed based on the literature review, and according to the Technology Acceptance Model. Variables with asterisk and in *italics* were adapted from Park (2009), the others were adapted from Hu and O'Brien (2016). However, each variable was modified by the authors to make it possible to measure the perceived role of AI in online shopping.

For data collection, a questionnaire made up of 26 questions (variables) was used (Table no. 1). Additionally, six demographics variables - gender, education, age, occupation, place of residence and internet subscription - were also included in the survey. All measurement instruments were listed in Table no. 1 but the demographics variables were measured on a seven-point Likert-scale ranging from strongly disagree (1) strongly agree (7).

In the very first section of the questionnaire, respondents were provided with a detailed explanation of AI-powered webshops and shopping apps, which are online stores where shopping is supported by artificial intelligence. AI-powered webshops present personalised product/service offerings based on previous search patterns and purchases that we made before, and automatically display products that AI chooses for us. Also, AI offers similar products to those that were originally viewed but were not available in the right size (product recommendation based on visual similarity). Another typical sign of an AI-powered webshop is that when the customer is leaving the web store, AI warns about the products left in the cart, to complete the purchase. AI-powered webshops often use chatbots, i.e. a virtual assistant is available if the customer has any questions, and visual (image-based) search is also possible: after uploading a product picture, AI recommends the most similar ones to that. Virtual changing rooms, voice recognition and automatic search completion are also available in AI-powered webshops such as Amazon, e-Bay, Alibaba, AliExpress, GearBest, eMAG.hu, PCland.hu, Ecipo, Bonprix, Answear, Reserved, Fashiondays, Fashionup, Spartoo, Orsay, to mention just a few.

**Table no. 1. Constructs and measurement instruments**

| Construct | Definition | Measurement Instruments |
|---|---|---|
| **Perceived Usefulness (PU)** | The degree to which a consumer believes that AI used in online shopping would make his or her purchases more effective. | PU1. The use of AI in retail (shopping ads and webshops) allows me to find the best deals.<br>PU2. The use of AI in retail enhances my effectiveness in purchasing.<br>PU3. The use of AI in retail is useful to me.<br>*PU4 The use of AI in retail saves time for me. ** |
| **Perceived Ease of Use (PEU)** | The degree to which a consumer believes that using AI in webshops will be free of effort. | PEU1. AI-powered shopping apps and webshops are easy to use.<br>PEU2. Shopping does not require a lot of my mental efforts if supported by AI (alternatives are offered by AI).<br>PEU3. Shopping is not so complicated if AI offers products to me.<br>*PEU4 Learning how to use AI-powered shopping apps and webshops is easy for me. **<br>*PEU5 It is easy to become skilful at using AI-powered shopping apps and webshops** |
| **Experience (EXP)** | The consumers' knowledge about and the experience with purchasing in an AI-powered webshop. | EXP1. I'm experienced in online shopping.<br>EXP2. I have already used AI-powered applications (chatbots, etc.) |
| **Trust (TRUST)** | The subjective probability with which people believe that AI works for their best interest. | T1. I am convinced that AI in retail is used to provide customers with the best offerings.<br>T2. I trust in apps and webshops that use AI. |
| **Subjective Norm (SN)** | The degree to which a consumer perceives that most people who are important to him or her think he or she should or should not make purchases in AI-powered webshops. | SN1. People who influence my behaviour would prefer me to use AI-powered shopping apps and webshops.<br>SN2. *I like using AI-powered webshops and shopping apps based on the similarity of my values and the social values underlying its use. ** |
| **Task Relevance (TR)** | The degree to which a consumer believes that AI-powered webshops are applicable to his or her shopping task. | TR1 I think AI can be used effectively in webshops and shopping apps. |
| **Compen-sation (COMP)** | The degree to which a consumer believes that he or she has the ability to make purchases in AI-powered webshops. | I would prefer AI-powered shopping apps and webshops…<br>C1. if there was no one around to visit physical shops/shopping malls with.<br>C2. if I had less time.<br>C3. if I had a built-in help facility for assistance when needed. |
| **Perceived Quality PQ** | The degree of how good a consumer perceives the quality of a product in AI-powered webshops. | PQ1 AI finds/offers better products for me than I could. |

| Construct | Definition | Measurement Instruments |
|---|---|---|
| **Perceived Enjoyment PE** | The extent to which shopping in AI-powered webshops is perceived to be enjoyable. | PE1 Shopping is more fun, enjoyable when AI helps me to find the best-suited products. |
| **Attitude ATT** | The consumer's attitude towards shopping in AI-powered webshops. | ATT1 Shopping in a webshop/shopping app that is powered by AI is a good idea<br>ATT2 Shopping in a webshop/shopping app that is powered by AI is a wise idea<br>ATT3 I am positive towards webshop/shopping app that is powered by AI |
| **Behavioural Intention BI** | A consumer's behavioural intention to do the shopping in AI-powered webshops. | BI1 I intend to visit webshops and to use shopping apps that are powered by AI more frequently.<br>BI2 I'm willing to spend more on products offered by webshops and apps powered by AI |

*Sources: Adapted from Hu and O'Brien, 2016; *Park, 2009.*

An online survey in Google Form was conducted to collect data in July and August 2020 in Hungary. Because of the Theory Acceptance Model, previous online shopping experience with AI-powered webshops was the one and only eligibility criterion for respondents to participate in this study. Convenience sampling method was used to reach the maximum number of respondents. Data was migrated from Google Form to MS Excel, SPSS 24 and AMOS, and was checked for coding accuracy. The database was complete and contained no missing data. Descriptive statistical analyses were done in SPSS. AMOS was employed to test the hypotheses and the theoretical model by structural equation modelling (SEM).

## 2.2. Research questions and hypotheses

Based on the literature review, this study aims to address the following research questions respectively:

- R1: Can the technology acceptance model (TAM) be used for investigating consumer acceptance of the use of artificial intelligence in online shopping?

- R2: If so, what are the key factors influencing behavioural intention to visit AI-powered webshops and apps?

Based on the Technology Acceptance Model, an initial theoretical model was developed (Figure no. 2). The arrows that link constructs (latent variables such as COMP, EXP, TRUST, SN, PEU, PU, ATT, BI) represent hypothesised causal relationships (hypotheses) in the direction of arrows. One of the objectives of this study is to test those hypotheses. Error terms for all observed indicators are indicated by e1 to e35, respectively.

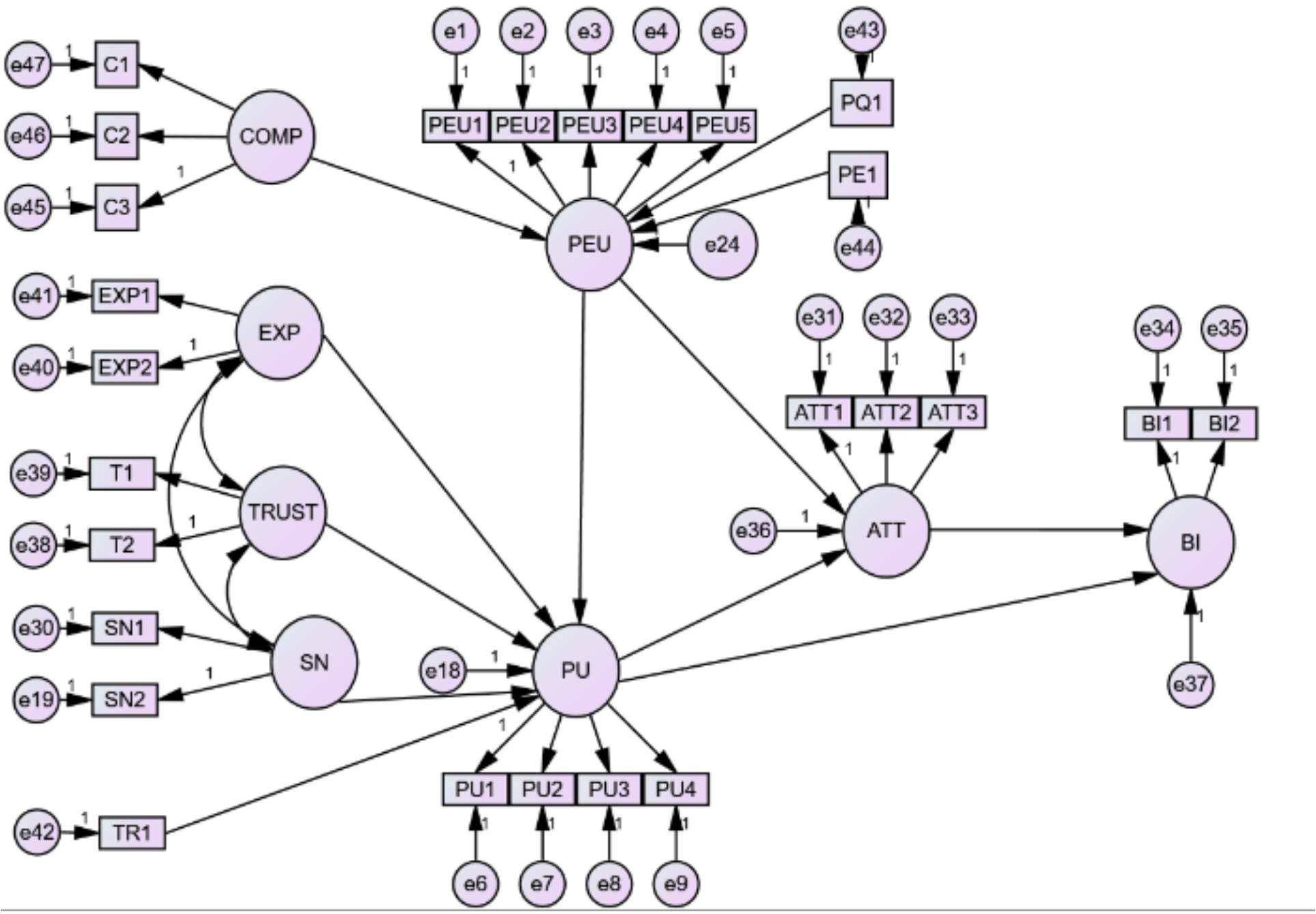

**Figure no. 2. The initial theoretical model**

## 2.3. The sample

A sample size of 200 is an appropriate minimum for SEM in AMOS (Marsh, Balla, and MacDonald, 1988), and a minimum of 10-20 subjects per parameter estimates in the model are optimal (Schumacker and Lomax, 2010). Therefore, the ideal sample size is between 380 and 760, considering the number of parameter estimates (38) in the initial model. The actual sample size of 439 respondents fits into this category.

Of the sample of 439 respondents, 62.2% were female, 37.8% male. Their average age was 32.2 years. 60,8% of the respondents had tertiary education, 38% had secondary education, and 1.2% had primary education. Most respondents resided in county seats (47.6%); the rest lived in other towns/cities (24.1%), villages (17.8%) and the capital (10.5%). Most respondents were subscribed to both mobile and wired internet services (84.3%), while 7.7% had only mobile internet, and 7.1% had only wired internet services. Only 0.9% of respondents had not got any subscription to internet services (wired or mobile). There is no data available on the distribution of the e-shoppers in Hungary, therefore, it is impossible to tell if this sample reflects the characteristics of the e-shoppers' population in Hungary.

## 3. Results and discussion

The initial model (Figure no. 2), which proved to be too complex and did not fit the current data (CMIN/DF=7.72; p=.00; GFI=,693; CFI=.723; RMSEA=.124; HOELTER 0.5= 65), was absolutely rejected. Therefore, it was not appropriate to interpret any individual parameter estimates, and further model modifications were required to obtain a better-fitting model. Respecification of the initial model led to a nested model that fitted well and is discussed further. During the respecification, the alternative model approach was used (Malkanthie, 2015). To test the model, the same data set was used. Several modified models were developed, and out of the theoretically justifiable models, the model with the best data fit was selected (Figure no. 3) as suggested by Mueller and Hancock (2008).

The respecification process was started with testing the measurement model by a series of Principal Component Analysis (PCA). Variables with factor loadings under 0.7 were deleted. A rule of thumb in confirmatory factor analysis suggests that variables with factor loadings under |0.7| must be dropped (Malkanthie, 2015). As a result, only one external variable, which is related to trust (T2), remained in the model (Table no. 2). Perceived Usefulness (PU) was measured by three variables (PU1, PU2 and PU3), whereas Perceived Ease of Use was made up of two variables (PEU2 and PEU3), and Behavioural Intention became unidimensional (B1). The attitude was composed of three variables (ATT1, ATT2 and ATT3). The nested model, which is theoretically consistent with the research goals, contains eight hypotheses:

- **H1:** Attitude has a positive effect on behavioural intention.
- **H2:** Perceived usefulness positively affects behavioural intention.
- **H3:** Perceived usefulness has a positive effect on attitude.
- **H4:** Perceived ease of use positively influences attitude.
- **H5:** Perceived ease of use positively influences perceived usefulness.
- **H6:** Perceived ease of use has a positive impact on trust.
- **H7:** Trust has a positive effect on perceived usefulness.
- **H8:** Trust positively influences attitude.

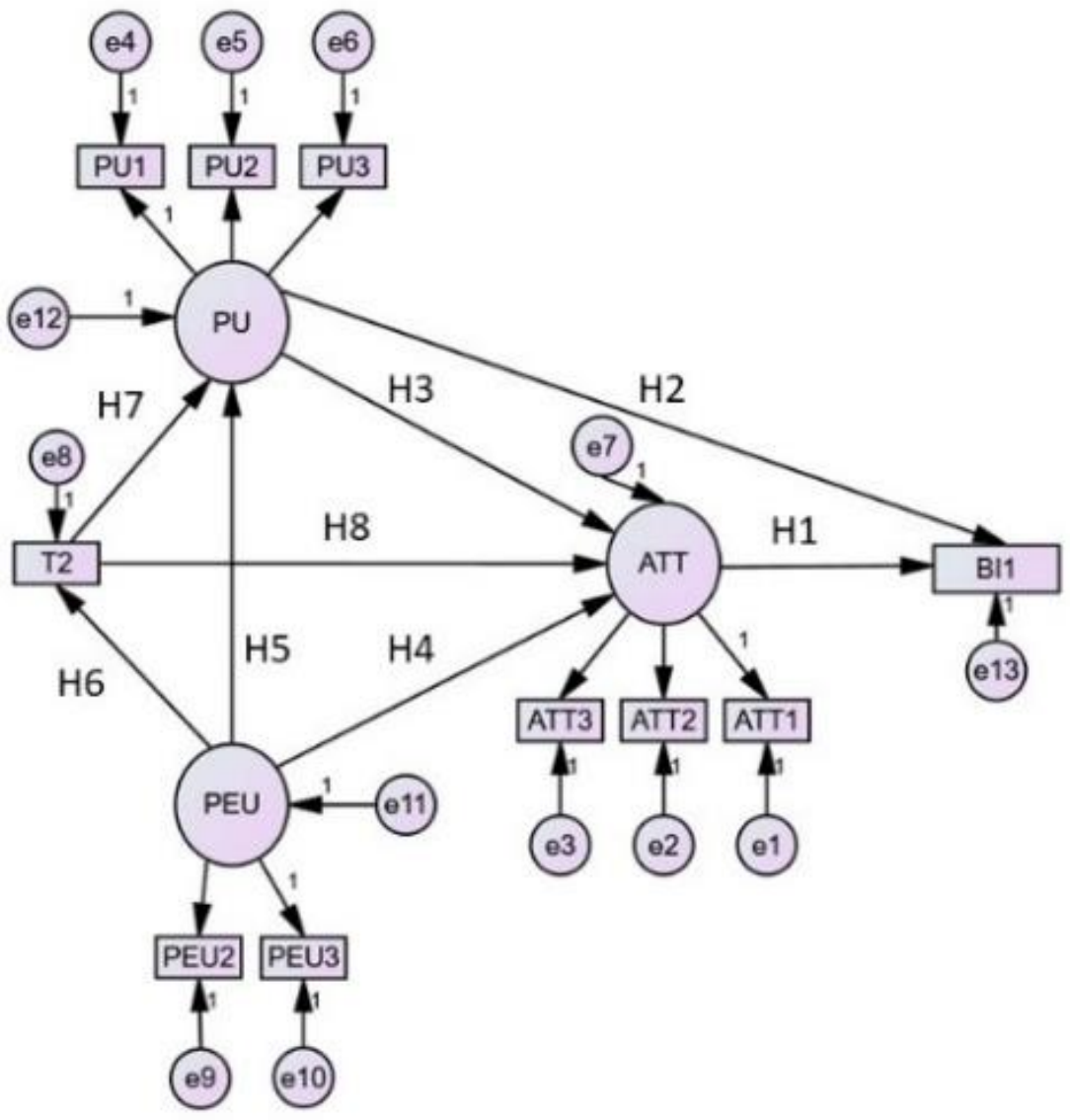

**Figure no. 3. The nested model**

### 3.1. Validity

To investigate the extent to which a set of items reflect the theoretical latent-construct they are designed to measure, both convergent and discriminant validity were checked. Convergent validity suggests that the variables of a factor that are theoretically related are expected to correlate highly. According to the Fornell-Larcker criterion for convergent validity, the Average Variance Extracted (AVE) should be greater than 0.5. According to the Hair, et al. (1998) criteria, AVE should be greater than 0.5, standardised factor loading of all items should be above 0.5, and composite reliability should be above 0.7.

In the nested measurement model, each factor loading was above .84 (Table no. 2).

**Table no. 2. Summary of means, standard deviations, normality, validity and reliability measures**

| Cons-truct | Measurement Instrument | Mean | STD | Z Skew | Z Kurt | Loa-ding | α | AVE | CR |
|---|---|---|---|---|---|---|---|---|---|
| Perceived Usefulness | PU1. The use of AI in retail (shopping ads and webshops) allows me to find the best deals. | 4.68 | 1.53 | -4.06 | -1.40 | 0.85 | 0.91 | 0.76 | 0.91 |
| | PU2. The use of AI in retail enhances my effectiveness in purchasing. | 4.67 | 1.63 | -4.56 | -1.87 | 0.89 | | | |
| | PU3. The use of AI in retail is useful to me. | 4.73 | 1.69 | -4.16 | -2.70 | 0.89 | | | |
| Perceived Ease of Use | PEU2. Shopping does not require a lot of my mental efforts if supported by AI (alternatives are offered by AI). | 5.15 | 1.62 | -6.38 | -0.59 | 0.90 | 0.88 | 0.81 | 0.9 |
| | PEU3. Shopping is not so complicated if AI offers products to me. | 5.06 | 1.64 | -6.44 | -0.68 | 0.90 | | | |
| Trust | T2. I trust in apps and webshops that use AI. | 4.11 | 1.62 | -2.00 | -2.90 | 1.00 | 1 | n.a. | n.a. |
| Attitude | ATT1 Shopping in a webshop/shopping app that is powered by AI is a good idea | 5.02 | 1.63 | -4.99 | -1.58 | 0.90 | 0.9 | 0.79 | 0.92 |
| | ATT2 Shopping in a webshop/shopping app that is powered by AI is a wise idea | 4.23 | 1.62 | -1.60 | -2.39 | 0.86 | | | |
| | ATT3 I am positive towards webshop/shopping app that is powered by AI | 4.72 | 1.70 | -4.11 | -1.87 | 0.90 | | | |

| Cons-truct | Measurement Instrument | Mean | STD | Z Skew | Z Kurt | Loa-ding | α | AVE | CR |
|---|---|---|---|---|---|---|---|---|---|
| Behavioural Intention | BI1 I intend to visit webshops and use shopping apps that are powered by AI more frequently. | 3.35 | 1.78 | 2.13 | -3.93 | 1.0 | 1 | n.a. | n.a. |

Notes: STD=Standard Deviation, Z Skew=Z score for skewness, Z Kurt=Z score for Kurtosis, α=Cronbach's alpha, AVE=Average Variance Extracted, CR=Composite Reliability, N=439.

Moreover, all AVE scores were also well above the threshold level (AVE (ATT)=0.79; AVE (PU)=.76 and AVE (PEU)=0.81), and all CR scores exceeded 0.7 (CR (PU)=.91; CR (PEU)=0.90 and CR (ATT)=0.92). Therefore, the model meets both the Fornell-Larcker (1981) criterion and the Hair et al. (1998) criteria for convergent validity, so the internal consistency of the model is acceptable.

To assess discriminant validity, i.e. the extent to which a construct is truly distinct to other constructs, AVEs were compared with squared inter-construct correlations (SIC). AVE scores higher than SIC scores indicate that discriminant validity is acceptable (ATT AVE=0.79, SIC1=0.61 and SIC2=0.32; PU AVE=0.76, SIC1=0.40 and SIC2=0.61; PEU AVE=0.81, SIC1=0.40 and SIC2=0.32). Discriminant validity was also confirmed by investigating correlations among the constructs. Since there were no correlations above .85, which is a threshold limit of poor discriminant validity in structural equation modelling (David, 1998), results also confirmed adequate discriminant validity (PEU*T2=0.52; PEU*PU=0.64; PEU*ATT=0.57; PEU*BI1=0.43; T2*PU=0.73; T2*ATT=0.74; T2*BI1=0.53; PU*ATT=0.78; PU*BI1=0.64; ATT*BI1=0.66).

### 3.2. Reliability

To test the accuracy and consistency of the nested model, three reliability tests were used: Cronbach's alpha (α), the Average Variance Extracted index (AVE) and Composite Reliability (CR). The threshold value for an acceptable Cronbach's alpha is .70 (Cronbach, 1951). The measurement model is acceptable if all estimates are significant and above 0.5 or 0.7 ideally; AVEs for all constructs are above 0.5 (Forner and Larcker, 1981); and finally, CRs for all constructs are above 0.7 (Malkanthie, 2015). Table no. 2 shows that the calculated Cronbach's alphas of all constructs were at least .87 or higher, and the AVE scores were also higher than 0.76, as well as the CRs were above 0.9; therefore, the reliability of the measurement model is optimal.

### 3.3. Model fit

Absolute- and relative model fits were tested. Each absolute measure was significant and indicated a good fit. Although Chi-square statistics are sensitive to large sample size and assume a multivariate normal distribution (Kelloway, 1998), even those measures were acceptable. However, other model fit indexes are better to consider as criteria. Therefore, the goodness-of-fit index (GFI), the adjusted goodness-of-fit index (AGFI), the root mean squared error of approximation (RMSEA) and the standardised root mean squared residual (SRMR) were also examined. All of them indicated a good absolute model fit. (Absolute measures: Chi square=34.154 (DF=29); Probability level=0.23; CMIN/DF=1.18; GFI=0.98; AGFI=0.96; RMSEA=0.02; SRMR=0.04). As far as the relative model fit is concerned, TLI or NNFI, GFI, AGFI, NFI, IFI, CFI and Critical N (CN or HOELTER) were calculated. All but CN range from zero to one. Values exceeding .9 show an acceptable fit, above .95 a good fit (Bentler and Bonnet, 1980). CN (HOELTER), which favours large samples over small ones (Bollen, 1990), is an improved method for investigating model fit (Hoelter, 1983). CN should be above 200 to indicate a good model fit. (Relative measures: TLI/NNFI=0.98; GFI=0.98; AGFI=0.96; NFI=0.93; IFI=0.99; CFI=0.99 and HOELTER (CN)=546). The results of the absolute and relative model fit test confirmed that the structural model is acceptable and suitable for the analysis and interpretation of the parameter estimates. Therefore, it can be concluded that *the technology acceptance model is suitable for investigating consumer acceptance of the use of artificial intelligence in online shopping*, which is the answer to the first research question (R1).

### 3.4. Hypothesis testing and estimates

Because of the non-normality of the variables in the nested model, the asymptotically distribution-free (ADF) method was used to estimate parameters in AMOS. ADF calculates the asymptotically unbiased estimates of the chi-square goodness-of-fit test, the parameter estimates, and the standard errors. The limitation of ADF is that it needs a large sample size (Bian, 2012), which criterion was met in this study (N=439). Skewness and Kurtosis z-values of the variables were out of the range of the normal distribution that is -2 and +2 (George and Mallery, 2010). Moreover, the p values of the variables were significant (p=.000) in the Shapiro-Wilk and Kolmogorov-Smirnov tests, which also confirmed non-normality.

To address the second research question (R2) and to determine the key factors influencing behavioural intention to use AI-powered webshops and apps, hypotheses were tested in the structural model (Table no. 3).

**Table no. 3. Direct, indirect, total effects and hypothesis testing**

| Hypothesis | Relationship | P | St. direct eff. | St. indirect eff. | St. total eff. | Result |
|---|---|---|---|---|---|---|
| H1 | BI1 ← ATT | *** | 0.41 | 0.00 | 0.41 | accepted |
| H2 | BI1 ← PU | *** | 0.32 | 0.19 | 0.51 | accepted |
| H3 | ATT ← PU | *** | 0.48 | 0.00 | 0.48 | accepted |
| H4 | ATT←PEU | 0.1 | 0.09 | 0.48 | 0.57 | rejected |
| H5 | PU ← PEU | *** | 0.35 | 0.28 | 0.64 | accepted |
| H6 | T2 ← PEU | *** | 0.52 | 0.00 | 0.52 | accepted |
| H7 | PU ← T2 | *** | 0.55 | 0.00 | 0.55 | accepted |
| H8 | ATT ← T2 | *** | 0.35 | 0.26 | 0.61 | accepted |

The arrows linking constructs represent hypotheses in the direction of arrows in the nested model (Figure no. 3 and Figure no. 4). Asterisks signal statistically significant relations between constructs. Gamma estimates were calculated from exogenous construct to endogenous construct, and beta estimates between two endogenous constructs. Figure no. 4 shows the standardised estimates, loadings and residuals regarding the relationships between constructs and observed indicators. A hypothesis was accepted if the presence of a statistically significant relationship in the predicted direction was confirmed.

As Table no. 3 shows, all hypotheses were accepted except for H4. So, the present findings, except for the relationship between perceived ease of use and attitude, are consistent with the Technology Acceptance Model proposed by Davis (1986). Surprisingly, perceived ease of use (PEU) was found to have no direct, significant effect on attitude (ATT), which is not in agreement with the original TAM (H4 rejected). This discrepancy could be attributed to the fact that shopping is not too complicated in AI-powered webshops, and it does not require too much mental effort. However, this slightly unexpected result coincides with the findings of a previous research by Ha and Stoel (2008), who examined the effect of PEU on attitude towards online shopping.

In this study, with H5 and H6 accepted, perceived ease of use (PEU) was found to have a significant, direct, positive impact on both the perceived usefulness (PU) and trust (T2). It suggests that the easier it is for a consumer to use an AI-powered webshop, the higher level of customer trust and perceived usefulness can be expected. Consumers trust in AI-powered shopping apps and stores that are easy to use, and consider those that are too complicated less useful. Similar results were obtained by Ha and Stoel (2008), who focused on consumers' acceptance of e-shopping. Gefen, Karahanna and Straub (2003) also found that perceived ease of use positively affected the perceived usefulness of a B2C website and the trust in an e-vendor.

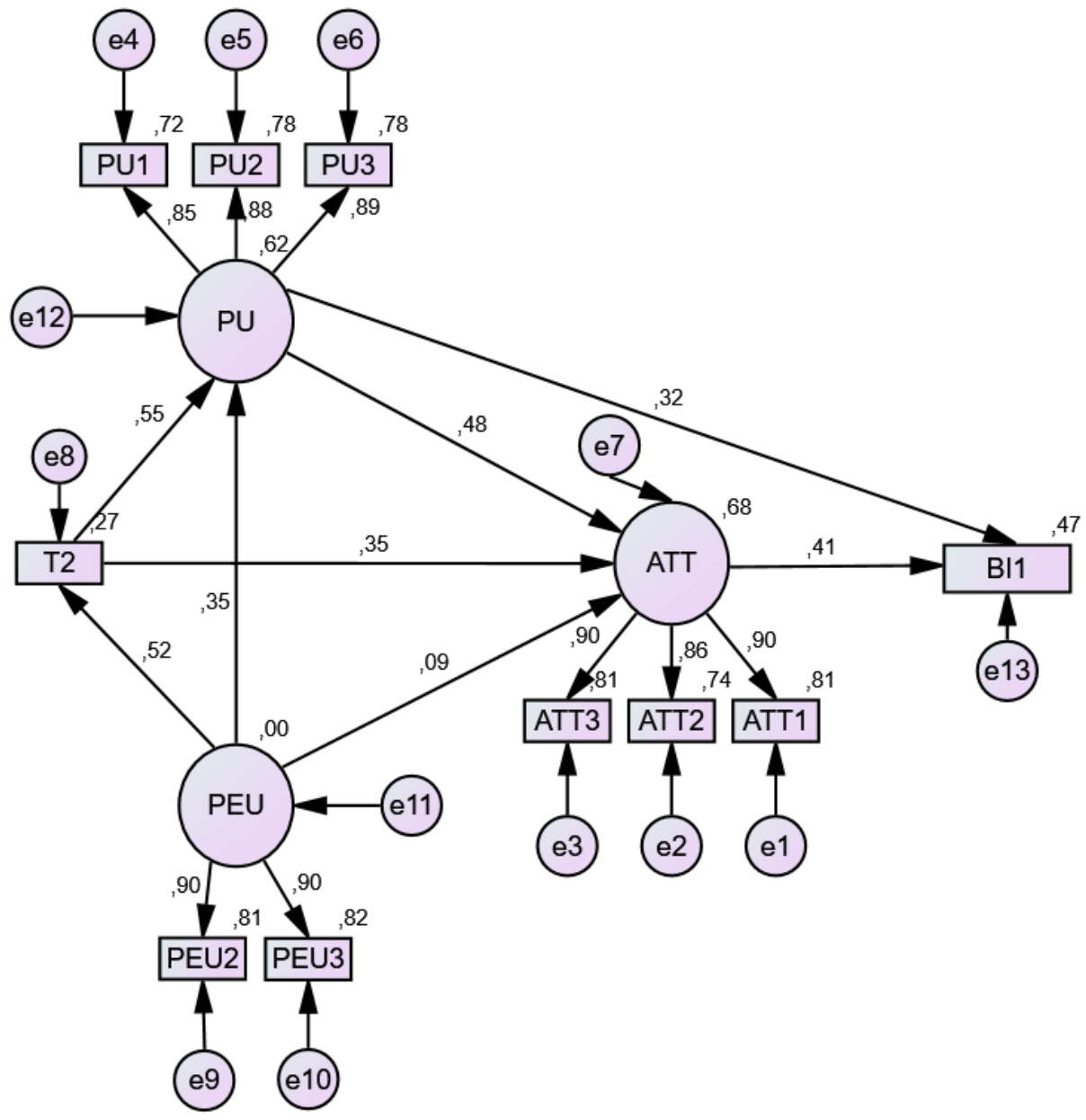

**Figure no. 4. Parameter estimates of the nested model**

*Trust in AI-powered webshops has a central role in forming attitudes and perceived usefulness.* Similar to what Gefen, Karahanna and Straub (2003), and Ha and Stoel (2008) found, trust directly influenced perceived usefulness (H7 accepted). Moreover, trust also impacted attitude (H8 accepted), in line with the research findings of Ha and Stoel (2008).

The strongest direct effect was found between trust and perceived usefulness (H7 accepted). It suggests that the more we trust in Artificial Intelligence during the online shopping journey, the more likely it is that we consider AI-powered apps and webshops useful. Besides, a higher level of trust forms a more positive attitude towards shopping in such webshops. Perceived usefulness has a central role in this model as it (PU) significantly impacted attitude (H3 accepted) and behavioural intention (H2 accepted). The more useful we find the use of artificial intelligence in online shopping believing that it allows us to grab the best deals, the more likely we are to consider it a wise decision to do the shopping in AI-powered webshops and apps more frequently. Not surprisingly, attitude towards AI-powered webshops and apps was found to have a strong, significant, positive direct impact on behavioural intention (H1 accepted). It suggests that forming consumers' attitude plays a vital role in increasing the traffic of AI-powered webshops and apps (Figure no. 4).

Although there was no significant direct relationship between perceived ease of use and attitude, the indirect effect of PEU on attitude (PEU->ATT=0.48) was quite strong, similar to its indirect impact on behavioural intention (PEU->BI1=0.43). Also, trust was found to indirectly influence behavioural intention (T2->BI1=0.42). It suggests that if shopping requires much mental effort and seems to be complicated in AI-powered webshops and apps, consumers tend to form stronger negative attitudes towards them and also tend to trust them less, which will result in weaker consumer intention to visit such webshops.

In the nested model perceived usefulness had the highest total effect on behavioural intention. *Therefore, AI-powered webshops and apps are advised to increase the level of perceived usefulness to succeed* by enabling customers to maximise purchase effectiveness to grab the best deals, i.e. the ideal product with the highest utility.

## Conclusions

This research extends our knowledge of consumer acceptance of the use of artificial intelligence in online shopping in many aspects. The widely used technology acceptance model (TAM) was proved to be suitable for investigating consumer acceptance of the use of artificial intelligence in online shopping.

As expected, it was confirmed in the nested model that the key factors influencing consumer' behavioural intention to use AI-powered webshops and apps are trust, perceived usefulness, perceived ease of use and attitudes. In contrast to the original TAM (Davis, 1986), the direct relationship between perceived ease of use and attitudes was insignificant. Nevertheless, it does not mean that user-friendliness of a webshop is not crucial as perceived ease of use indirectly affects attitude and the behavioural intention. Instead, user-friendliness and flawless operation of an artificial intelligence-powered website are the prerequisites for market success.

Building trust has a central role in consumer acceptance of the use of artificial intelligence in online shopping. If consumers do not trust in an AI-powered webshop/app, they tend to consider it less useful and form a negative attitude towards it, which will result in less online traffic. Also, AI must provide online consumers with tailor-made offerings to grab the best deals, i.e. products with the highest value; and it is expected to shorten the product search time to enhance shopping effectiveness. Not surprisingly, the favourable attitude towards AI-powered webshops leads to more frequent online traffic in such electronic stores.

Considering the strong positive impact of the recent COVID-19 crisis on e-commerce, the use of artificial intelligence in online shopping is expected to expand further. According to Bloomberg (2020) the pandemic lockdowns have a dual effect on consumer behaviour on the development of AI. Nowadays, it is more important than ever to create a personalised customer journey, to meet customers' demand and to provide a greater online shopping experience. In these efforts, artificial intelligence can be a very effective tool, which was confirmed by the research findings of this paper.

This study has several practical applications. It is useful for webshop owners and online marketing managers to understand how consumers adapt to the new technology, i.e. the use of artificial intelligence in online shopping. It is also beneficial to academics and researchers who are interested in the adaptation of the Technology Acceptance Model in online shopping. Those who are interested in the role of trust in consumer choices in the online environment will also benefit from this study.

As far as the future research directions are concerned, it would be advisable to repeat this study in a multi-cultural context. It might also be useful to test the model of the Technology Readiness Index proposed by Parasuraman (2000) and to compare the results presented here with the new findings.

**Acknowledgements**

“The described article/presentation/study was carried out as part of the EFOP-3.6.1-16-2016-00011 “Younger and Renewing University – Innovative Knowledge City – institutional development of the University of Miskolc aiming at intelligent specialisation” project implemented in the framework of the Szechenyi 2020 program. The realization of this project is supported by the European Union, co-financed by the European Social Fund.”